# Gigantic magnetic field induced polarization and magnetoelectric coupling in a ferrimagnetic oxide $CaBaCo_4O_7$

V. Caignaert[1], A. Maignan[1]*, K. Singh[1,5], Ch. Simon[1], V. Pralong[1], B. Raveau[1], J.F. Mitchell[2]

H. Zheng[2], A. Huq[3], and L. Chapon[4]

[1] Laboratoire CRISMAT, UMR 6508 CNRS/ENSICAEN, 6 bd du Maréchal Juin
F-14050 CAEN Cedex 4 – France.

[2] Argonne National Laboratory MSD 223 9700 S. Cass Avenue Argonne, IL 60439, USA

[3] Neutron Scattering Science Division, Oak Ridge National Laboratory, Oak Ridge, Tennessee, 37831, USA

[4] Institut Laue-Langevin 6, rue Jules Horowitz - BP 156 F- 38042 Grenoble Cedex 9, France

## Abstract

The single crystal study of $CaBaCo_4O_7$, a non collinear ferrimagnet ($T_C$=64K), with a polar orthorhombic space group ($Pbn2_1$) between 4 K and 293 K, shows the appearance below $T_C$ of a large electric polarization along its $\vec{c}$ axis, reaching 17 mC.m$^{-2}$ at 10K. At 62.5 K, a magnetic field driven giant variation of polarization, P(9T) - P(0T) = 8 mC/m$^2$, is observed. Moreover, the present magnetoelectric measurements are fully consistent with the m'm2' magnetic point group, strongly supporting that this oxide is also ferrotoroidic. This ferrimagnetic oxide, which belongs to the "114" structural family, opens an avenue for the search of new magnetoelectrics.

* Antoine Maignan
Laboratoire CRISMAT, ENSICAEN/CNRS, 6 boulevard du Maréchal Juin, 14050
Caen cedex 4 - France
antoine.maignan@ensicaen.fr
Tel: 02.31.45.26.04
Fax: 02.31.95.16.00

Numerous investigations of multiferroics have shown that two physical characteristics of these materials are of great importance in view of technological applications, the magnetoelectric coupling and the electric polarization, which should be as high as possible [1-2]. Based on these two prerequisites, improper ferroelectrics, where ferroelectricity originates from a particular magnetic order, are challenging for discovering new performances and understanding multiferroism. Significant coupling between magnetism and ferroelectricity was observed in a rather large number of improper ferroelectrics, but the coefficients of the tensor characterizing the linear magnetoelectric effect $\alpha_{ij}$ were rarely measured. The highest value of $\alpha$ that has been reported to date is close to 20000 ps/m for $Ba_{0.5}Sr_{1.5}Zn_2(Fe_{0.92}Al_{0.08})_{12}O_{22}$ [3]. Unfortunately, it is also observed that the magnetically induced polarization of these magnetoelectric ferroelectrics remains rather low, generally smaller than 100 $\mu Cm^{-2}$, as shown for example for $TbMnO_3$ [4], $MnWO_4$ [5], $TbMn_2O_5$ [6], $Ni_3V_2O_8$ [7]. Recently, the multiferroic $GdMn_2O_5$ was shown to exhibit giant ferroelectricity [8-9], with ~3600 $\mu C/m^2$, the largest observed value for improper ferroelectrics.

The "114" $CaBaCo_4O_7$ cobaltite [10-12] exhibits a pure tetrahedral framework [10-11] where the $CoO_4$ tetrahedra are three-dimensionally interconnected and form a geometrically frustrated network (Fig. 1). The magnetic structure of this oxide shows that, similarly to several improper ferroelectrics which are antiferromagnets, the cobalt spins are non collinear but differently from the latter, $CaBaCo_4O_7$ is ferrimagnetic, below $T_C$~64K, with $\vec{b}$ as easy axis. In contrast to many improper ferroelectrics, it crystallizes in a noncentrosymmetric space group $Pbn2_1$, in the whole temperature range, from 4 K to 293 K, with $\vec{c}$ as polar axis. Studies on polycrystalline samples have shown that below 64 K the magnetic ordering induces an additional polarization [12], suggesting the existence of improper ferroelectricity. In order to conclusively determine the nature of the magnetoelectric coupling and of the electric polarization in this phase, a single crystal study was necessary. Such a study is also motivated by the fact that the magnetic order in this oxide reduces the point group symmetry to m'm2', similarly to several magnetoelectric boracites [13-17], so that the existence of ferrotoroidicity can be predicted [12]. Here, we show that $CaBaCo_4O_7$ exhibits a high magnetoelectric coupling factor leading to a magnetic field driven gigantic change in the polarization near $T_C$. Neutron diffraction data reveal an abrupt structural change at $T_C$, and it is proposed that the electrical polarization has a magnetostrictive origin.

Millimeter size crystals were grown using the floating zone technique in a mirror furnace under air at 3.5 bars. Laue diffraction patterns showed that the single crystals exhibit the same characteristics as the polycrystalline samples with the space group $Pbn2_1$. Laue pattern were collected to obtain the geometric relation between crystal faces and the crystallographic axes. Finally platelet-like crystals were cut with the thinnest dimension (~1mm) corresponding to the $\vec{c}$ axis and with the largest faces (*xy* planes) reaching 5x5 mm.

Figure 2a shows the field cooled magnetization measured along the easy-axis ($\vec{b}$) upon warming at $10^{-2}$T. A sharp transition is evidenced with $T_C$=64 K, in good agreement with data previously reported for the polycrystalline precursor [10]. The temperature dependence of the dielectric permittivity (ε') was measured along the $\vec{c}$ and $\vec{b}$ directions (Fig. 2b and inset). Along $\vec{c}$, a clear peak is observed at $T_C$, supporting the possibility of a magnetoelectric coupling along $\vec{c}$ as predicted by the symmetry analysis [see ref. 12]. A second peak, with no corresponding anomaly on the M(T) curve was also observed at 69 K, implying a non-magnetic origin. In contrast, along $\vec{b}$ (easy-axis for magnetization) only a change of slope is observed at $T_C$ with a sharp drop of ε' at $T_C$. To test the origin of the anomaly above $T_C$, specific heat measurements were made (on cooling), without and with an applied external magnetic field of 2 T, using a larger crystal (Fig. 2c). The peak at $T_C$ with $\mu_0H$=0 becomes broader and shifts towards higher T within $\mu_0H$=2 T. It corresponds to the ferrimagnetic ordering. For H=0, a second peak is detected at 69 K, which does not change in fields measured up to 2 T (Fig. 2c, inset). This strongly supports the lack of magnetic origin for this small, high temperature peak on the ε'(T) curve (Fig. 2b).

The presence of a well-defined dielectric peak along $\vec{c}$ at $T_C$ motivated polarization measurements. For that purpose, a thin platelet with contacts on the largest *xy* faces was cut from the crystal used for the M(T) in order to apply a small electric field along $\vec{c}$ (E=1.1 kV/cm) during the cooling from 80 K to 8 K. At 8 K, *E* was removed and *P* was measured upon warming at 0.5K/min (Fig. 2d). A gigantic variation of the polarization is evidenced, between 10K and 80 K. Also, the sharp transition at $T_C$ towards ***P***=0 demonstrates the improper origin of the electric polarization in the magnetically ordered phase. However this polarization cannot be reversed completely by changing the polarity of a poling electric field (E=±14kV/cm). Moreover it should be pointed out that an electric polarization is observed below $T_C$ even in a zero poling electric field. Above $T_C$, the polarization remains constant (*i.e.*, the pyroelectric current is null) but its variation cannot be measured above ~150K due to the high value of the dielectric losses.

To demonstrate the existence of a magnetoelectric (ME) coupling, *P* measurements were performed under magnetic field. It must be mentioned that with our experimental set-up, no polarization measurements under both magnetic and electric field could be made. So, the coupling terms $\gamma_{xyz}$, linear in H and E, and allowed in the ordered magnetic state [12], could not be determined. Moreover, considering the anisotropic shape of the crystals, the *P* measurements were made only along the thinnest direction, $\vec{c}$, which is also the polar axis. In that geometry, upon application of an external magnetic field H, the induced polarization along $\vec{c}$ axis is given by $P_z=\alpha_{32}H_y + (\beta_{311}H_x^2+\beta_{322}H_y^2+\beta_{333}H_z^2)/2$ (eq.1), which $\alpha_{ij}$ is the coefficient of the linear magnetoelectric tensor and $\beta_{ijk}$ is the coefficient of the bilinear magnetoelectric susceptibility tensor. Applying H along $\vec{b}$, the formula reduces to: $P_z=\alpha_{32}H_y + \beta_{322}H_y^2/2$ (eq.2), from which the two coefficients $\alpha_{32}$ and $\beta_{322}$ can be extracted below $T_C$ while above $T_C$, only $\beta_{322}$ is allowed. As aforementioned, H was applied along the easy-axis $\vec{b}$ in order to be perpendicular to *E* ($\vec{c}$ axis). The induced polarization $P_z$ was recorded from 0 to 9 T at several temperatures between 10 K and 75 K, i.e. below and above $T_C$. An increase of the polarization with the applied field is observed whatever the temperature in this range (Fig.3). The values of $\alpha_{32}$ and $\beta_{322}$ coefficients as a function of T, obtained by fitting the $P_z(H_y)$ curves (inset of Fig. 4) between -3T and 3T with eq.2, are given in Fig. 4.

From these curves, it is clear that the temperature dependence of $\alpha_{32}$ goes through a maximum just below $T_C$, as observed for the Ni-Cl or Co-I boracites [13, 15]. At 60 K, the value of the coefficient of the tensor $\alpha_{32}$ for the linear ME coupling, obtained by fitting the $P_z(H_y)_{T=60K}$ curve as shown in Fig. 4, reaches $\alpha_{32}$=764 ps/m value in SI unit. This compares with the value $\alpha_{xy}$=730 ps/m reported at 1.5 K for a TbPO$_4$ single crystal [18]. As expected for the paramagnetic state, $\alpha_{32}$ is found to be close to zero for T>$T_C$. The bilinear coefficient $\beta_{322}$ is negative below $T_C$ and positive above. Near the magnetic transition $\beta_{322}$ decreases abruptly towards negative values, changes sign at the transition and decreases again with temperature (Fig.4). A similar behavior has been previously measured in Ni-Cl boracites [13]. Additional measurements were also made to verify the predictions coming from the m'm2' point group. Under application of H along $\vec{c}$ (H$_z$), the induced polarization P$_z$ should only depend on $\beta_{333}$ with $P_z = \beta_{333}H_z^2/2$, since the linear magnetoelectric coefficient $\alpha_{33}$ is expected to be equal to zero by symmetry. At 10K, the $P_z(H_z)$ curve leads to $\beta_{333}$=18.4(2) as/A and $\alpha_{33}$=0.07(11) ps/m, i.e. $\alpha_{33}\approx0$, as expected for the point group m'm2'. Thus, the present ME$_H$ measurements with H along $\vec{b}$ and $\vec{c}$ confirm the magnetic point group m'm2' for CaBaCo$_4$O$_7$. As toroidization is allowed in this point group [12], CaBaCo$_4$O$_7$ may also be ferrotoroidic. In that respect, the divergence of $\alpha_{32}$(T) near $T_C$ is an indirect method to probe the existence of a toroidal moment [19]. The shape of the $\alpha_{32}$(T) curve for $CaBaCo_4O_7$ is consistent with the theory (for a review see the references in [18] and [19]). In addition, at T=10 K, a butterfly loop in the P(H) curve is observed (inset of Fig. 3) with a characteristic symmetric minimum, corresponding with the coercive magnetic field (~0.6 T) and consistent with a spontaneous magnetization as for $LiCoPO_4$ [20] and Ni-I boracite [21].

As shown in Fig. 3, the largest ME effects are achieved close to $T_C$. This motivated the measurements of H-dependent M, ε' and ΔP at 65K (Fig. 5). A large magnetodielectric effect of 80% is found in only 1T (ε'(H), Fig. 5b) together with a large magnetoelectric response ΔP(H) (Fig. 3). This can be compared to the derivative of the magnetization with respect to magnetic field (inset of Fig. 5a), showing a maximum at ~1-1.5 T. A metamagnetic transition occurs from a paramagnetic state below $\mu_0$H~1T towards an ordered magnetic state above that value. This transition is reflected by the derivative curve of ΔP(H) (right inset, fig. 5b) and confirmed by the value of magnetoelectric coefficients (linear and bilinear), calculated from the ΔP(H) curve. The $\beta_{322}$ coefficient is positive below 1 T as in the paramagnetic state with a value 2.0(3) fs/A. In contrast the $\beta_{322}$ coefficient is negative above 1 T as is observed in the ferrimagnetic state below $T_C$. It should be pointed out that the sign change of the bilinear coefficient, i.e. the metamagnetic transition, is also observed at 68 K (fig.3) but under a higher field, around 7 T.

Although the present measurements of pyroelectric current indicate that exceedingly high values of induced polarization are achieved in this non collinear ferrimagnet, the lack of evidence for *P* switching argues that this oxide is not a ferroelectric below $T_C$. Nevertheless, its magnetoelectric effects are remarkable. The variation of the polarization under magnetic field reaches values as high as ~ 8 mC/m$^2$ around $T_C$ between 62.5 and 65 K, which is nearly twice the largest variation among the known magnetoelectric compounds [8]. The present results indicate also that exceedingly high values of induced *P* in the magnetic state can be achieved in non collinear ferrimagnets. The values for $CaBaCo_4O_7$ are higher by almost three orders of magnitude than that recorded at 300K for the Z-type hexaferrite, $Sr_3Co_2Fe_{24}O_{41}$, ~25μC/m$^2$ [22] and by a factor of five compared to $CaMn_7O_{12}$ [9] or $GdMn_2O_5$ [8].

We now consider possible origins for the high values of Δ*P* and magnetoelectric coupling coefficients. Different mechanisms have been invoked to explain improper ferroelectricity in centrosymmetric materials: spin spiral magnetic order which breaks the inversion symmetry [23], and asymmetric exchange as in a chain of alternating magnetic ions with antiferromagnetic and ferromagnetic exchanges [24]. The case of some boracites and

CaBaCo$_4$O$_7$ differs from these examples as the latter compounds are polar by symmetry with a possible electric polarization in their paramagnetic states. They exhibit an extra polarization in their ordered magnetic state and a magnetoelectric coupling resulting from the m'm2' point group. However, $\Delta P$ and $\alpha$ values are much higher in the case of CaBaCo$_4$O$_7$ as compared to magnetoelectric boracites. Moreover, since the variation of the polarization and the values of the magnetoelectric coefficients are maximum near T$_C$, the magnetostriction may indeed play a major role. Our measurements by neutron diffraction on polycrystalline CaBaCo$_4$O$_7$, using the POWGEN diffractometer at Oak Ridge National Laboratory, revealed small but abrupt variations of the unit cell parameters to be detected at T$_C$. As shown in Fig.6, the variation of the cell parameter below T$_C$ follows the variation of both the polarization and the magnetization. This suggests that the increase of the polarization below T$_C$ is strongly linked to the magnetostriction. However the structure of CaBaCo$_4$O$_7$ is quite complex and the relatively large values of the estimated standard deviations on positional parameters preclude a quantitatively reliable calculation of the induced polarization. The sensitivity of the structural and electrical properties to the spin ordering is also reflected by the effect of an external magnetic field. As shown in Fig. 5a, at 65 K (i.e. just 1K above T$_C$) and for H>1 T, a metamagnetic transition from a paramagnetic to a ferrimagnetic state is induced. Thus, magnetic field application could control the change from the polar paramagnetic phase to the polar ferrimagnetic phase with a gigantic variation of the polarization.

This study of a CaBaCo$_4$O$_7$ single crystal demonstrates that its ferrimagnetic ordering induces a gigantic variation of its electric polarization near T$_C$, five times larger than the highest value reported for GdMn$_2$O$_5$ [8]. Moreover, its linear magnetoelectric effect is also one of the highest that has been reported up to now.

Remarkably, this phase which belongs to the "114" structural family exhibits a pure tetrahedral coordination of cobalt, differently from most of magnoelectrics where the magnetic cations are either in octahedral or pyramidal coordination. Several oxides of this "114" series such as YbBaCo$_4$O$_7$ [25], TmBaCo$_4$O$_7$ [cited in 25] and YBaCo$_4$O$_7$ [26-27] have been shown to exhibit, like CaBaCo$_4$O$_7$, a $k$=0 magnetic propagation vector which is a symmetry condition for the appearance of spontaneous magnetization and linear magnetoelectric effect. Thus, the "114" family offers a potential new route toward design of performing magnetoelectrics, and a challenge for understanding the role of the magnetostriction in the appearance of induced polarization and strong magnetoelectric coupling.

**Acknowledgment**: Work in the Materials Science Division of Argonne National Laboratory (single crystal growth, heat capacity measurements) is sponsored by the U.S. DOE, Office of Science, Office of Basic Energy Sciences, Materials Science and Engineering Division under contract No. DE-AC02-06CH211357. The Authors thank Laurence Hervé (CRISMAT) for crystal growth.

[5] current address: UGC-DAE Consortium for Scientific Research, University Campus, Khandwa Road, Indore, 452017, India

**Figure captions:**

Fig 1: Crystal structure of $CaBaCo_4O_7$.

Fig.2: a) FC Magnetization along the *b* axis; Cooling and magnetic measurement were carried out under 100 Oe; b) Dielectric permittivity along the *c* axis at 100 kHz, and along the *b* axis (inset); c) Specific heat; Inset of c shows enlargement around $T_C$ of specific heat measured in 0 T (blue dots), 1T (green down triangles) and 2 T (red up triangles); d) Variation of the polarization along the *c* axis after poling under $E_c$=1.1 kV/cm. All the measurements were performed upon heating except the specific heat measurement.

Fig.3: Magnetic field dependence of the polarization along the *c* axis at various temperatures. Magnetic field is applied along the *b* axis. Inset: polarization hysteresis loop measured at 10 K.

Fig. 4: Linear magnetoelectric coefficient $\alpha_{32}$ (dots) and bilinear magnetoelectric coefficient $\beta_{322}$ (squares) versus T in SI units (Doted lines are guided for eyes). Inset: Least-squares fit of $P_z(H_y)$ above (bottom) and below (top) $T_C$. For clarity the figure shows only the fit between 0 and 3T. At T=65K, the fit of $P_z(H_y)$ is performed between to -1T and +1T.

Fig. 5: Isothermal curves collected at 65 K a) Magnetization versus field with ***H**//**b***; b) Magnetodielectric effect at 100 kHz versus field with ***E**//**c*** and ***H**//**b***; Inset enlargements for $0T \leq \mu_0 H \leq 3T$ of (a) $\Delta M/dH = f(H)$, (b-left) $\varepsilon'(H)$ and (b-right) $\Delta P/dH = f(H)$.

Fig. 6: From top to bottom: ***a***, ***b*** and ***c*** cell parameters of polycrystalline $CaBaCo_4O_7$ versus T refined from neutron powder diffraction data.

**Figure 1:**

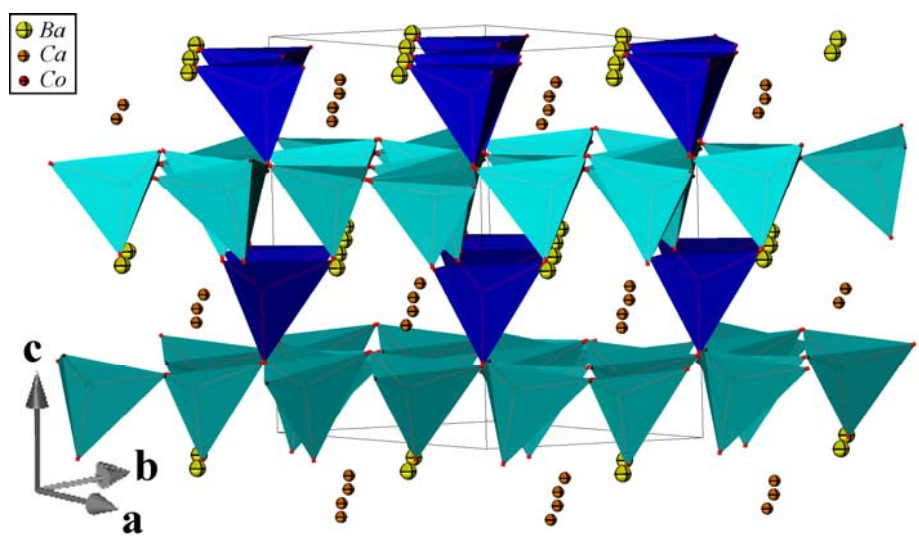

**Figure 2:**

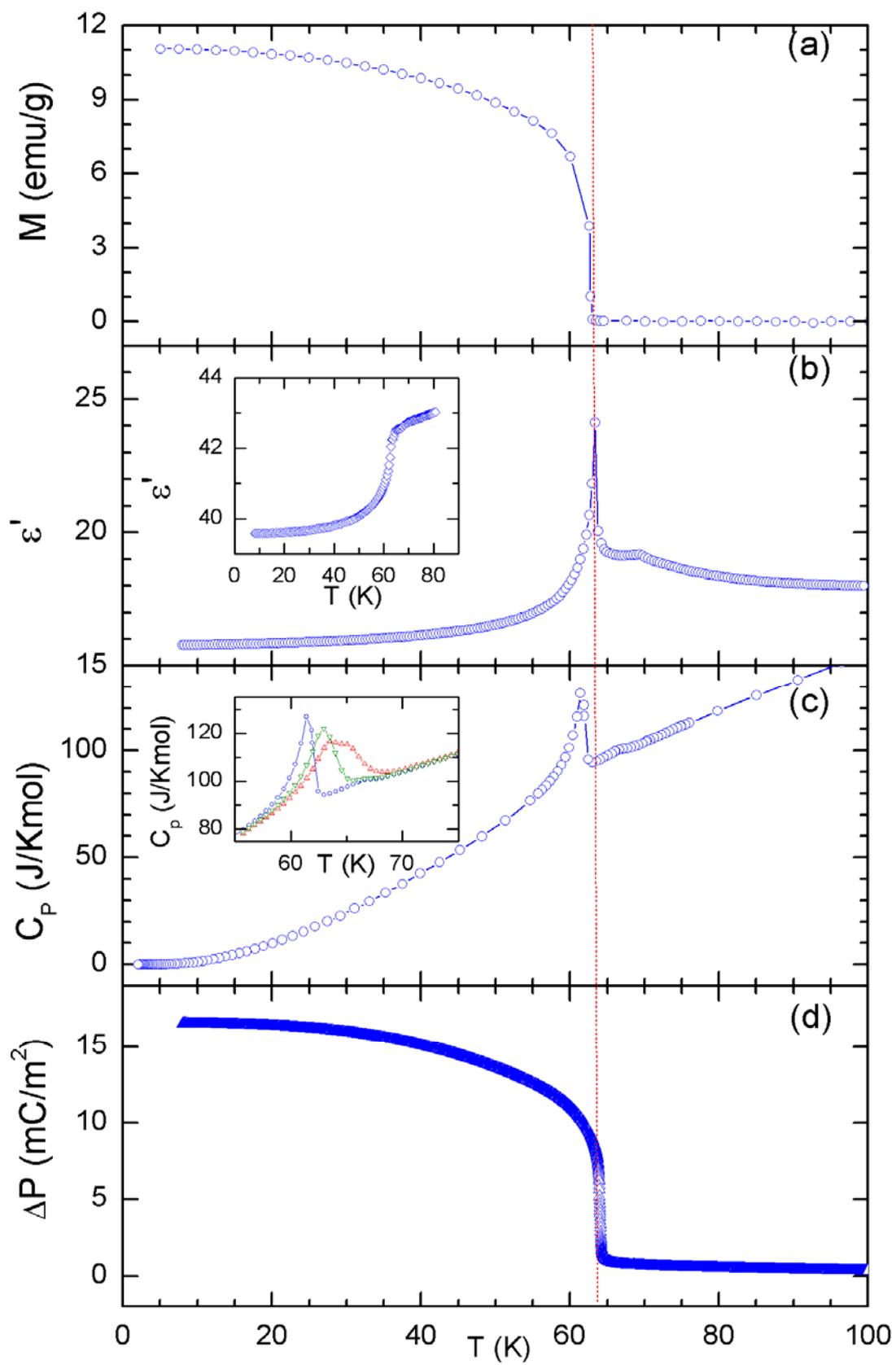

**Figure 3:**

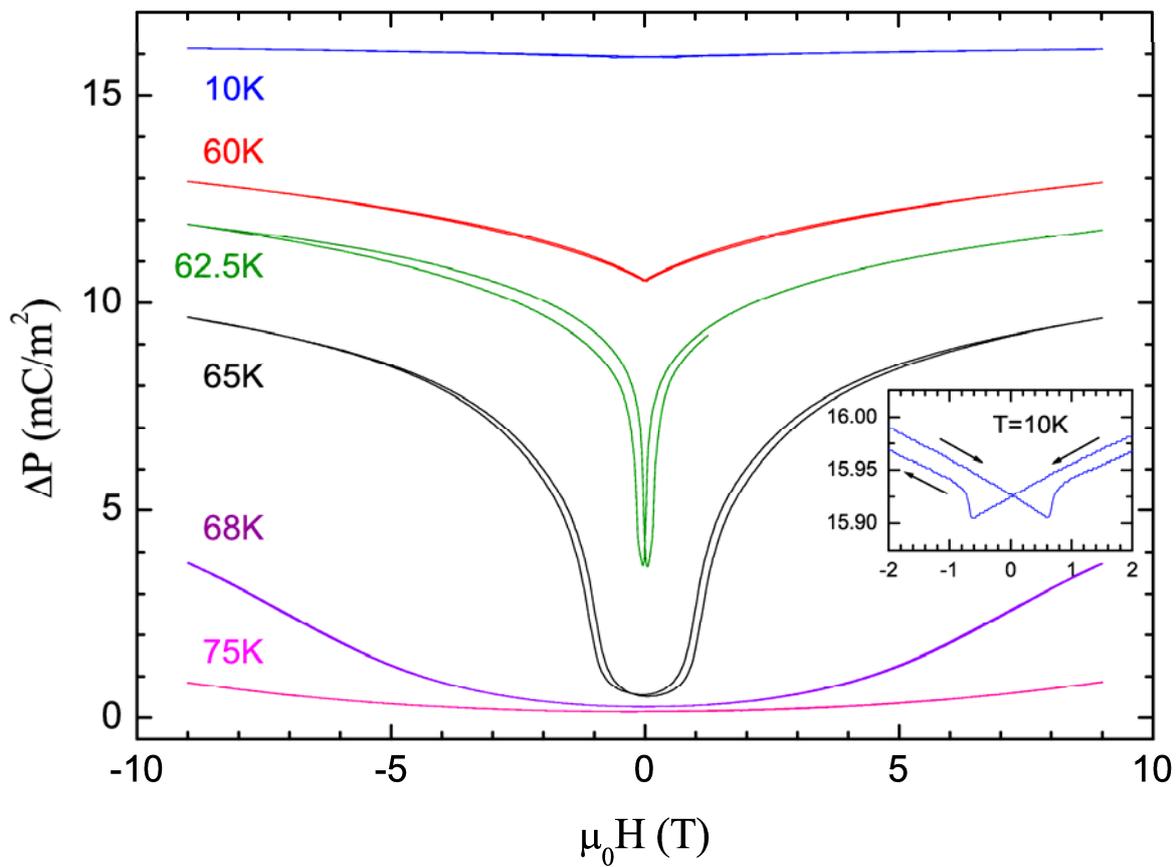

**Figure 4:**

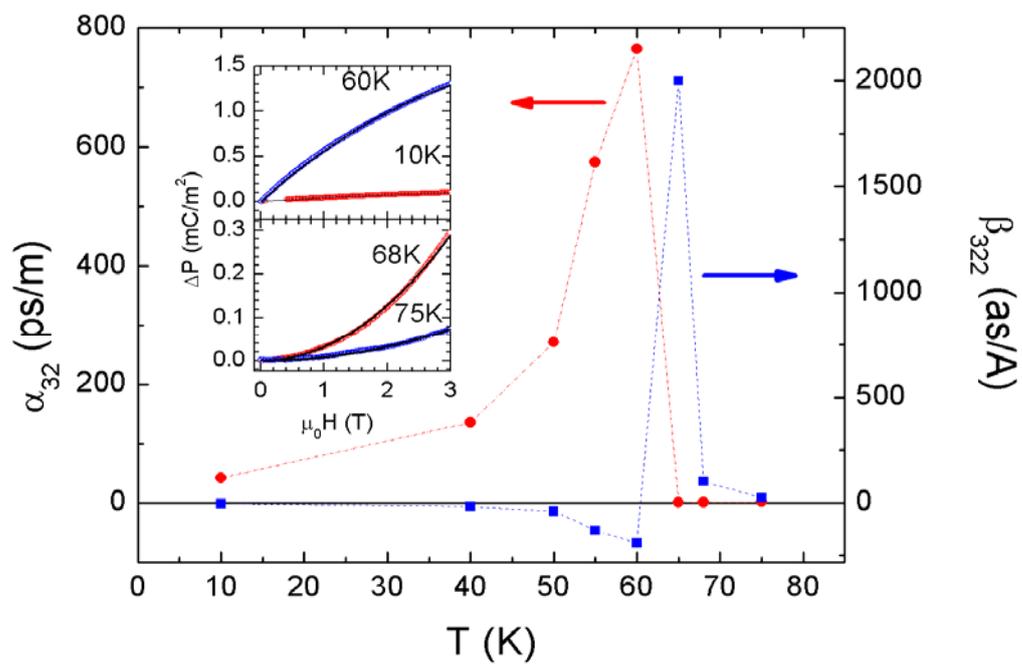

**Figure 5:**

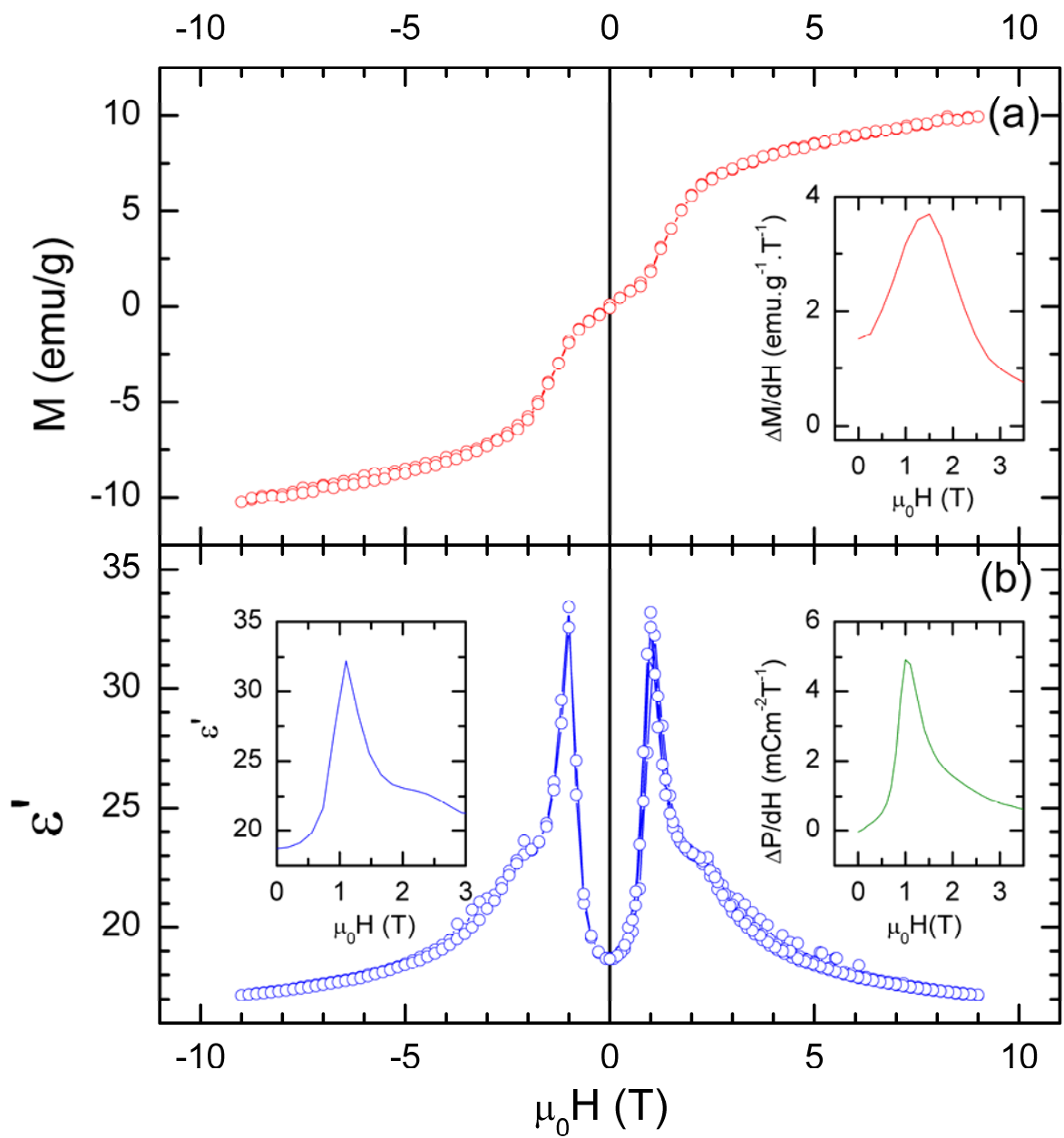

**Figure 6:**

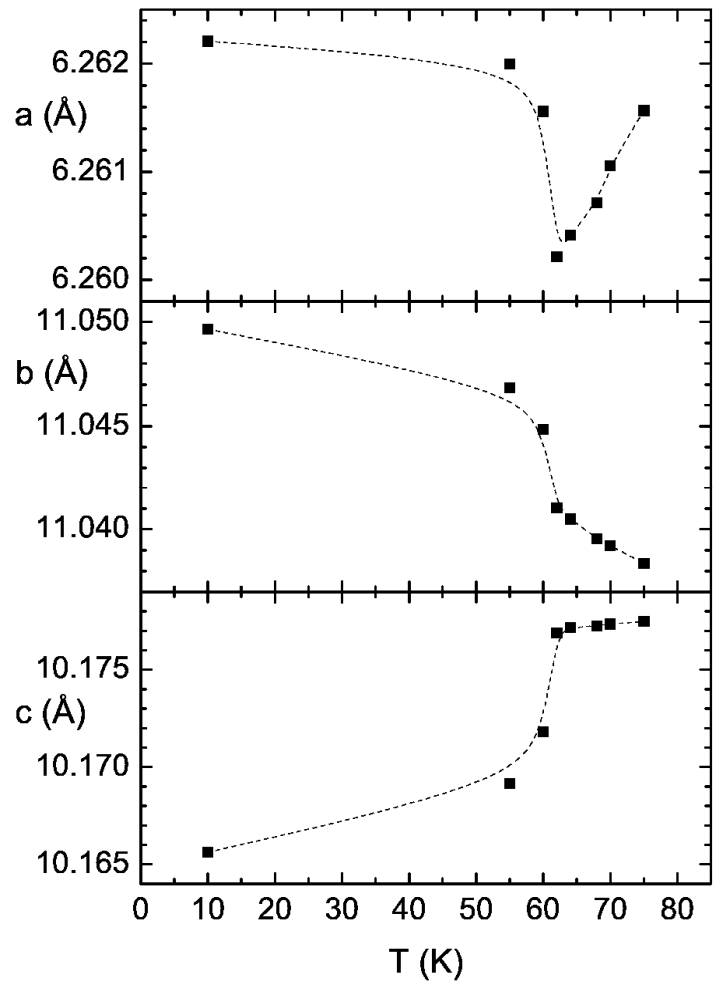